\documentstyle[prl,aps,twocolumn,floats]{revtex}
\input psfig.tex
\begin{document}
\twocolumn[\hsize\textwidth\columnwidth\hsize\csname
@twocolumnfalse\endcsname
\preprint{CERN-TH/96-98}
%\rightline{arch-ive/9707286}
\title{Structure formation with a self-tuning scalar field}
 
\author{Pedro G. Ferreira$^1$ and Michael Joyce$^2$}

\address{$^1$Center for Particle Astrophysics,301 Leconte Hall,
 University of California, Berkeley, CA 94720, USA.}
\address{$^2$School of Mathematics, Trinity College, Dublin 2,
Ireland.}
\date{July 1997}
\maketitle
\begin{abstract}
A scalar field with an exponential potential has 
the particular property that it is attracted into  
a solution in which its energy scales as the dominant component
(radiation or matter) of the Universe, contributing a fixed 
fraction of the total energy density.
We study the growth of perturbations in
a CDM dominated $\Omega=1$ universe with this extra field, with 
an initial flat spectrum of adiabatic fluctuations. 
The observational constraints from structure formation are 
satisfied as well, or better, 
than in other models, with a  contribution to the energy density 
from the scalar field $\Omega_\phi \sim 0.1$ which is small enough to
be consistent with entry into the attractor prior to nucleosynthesis. 
\\
\end{abstract}
]
%\end{abstract}
%\vskip0.5pc]

%\vspace*{-17.5cm}
%\hspace*{9cm} 
%\mbox{preprint/...., astro-ph/......}
%\newpage
\noindent

The simplest viable cosmology which follows from inflation,
a flat universe with pressureless matter and $5\%$  
baryonic dark matter, has been unable to fit both 
the cosmic background radiation (CBR) fluctuations
and measurements of mass fluctuations on  scales
of a few Megaparsecs. The paradigm of inflation 
is sufficiently compelling that there have been various 
attempts at modifying   this 
`standard cold dark matter' (sCDM) scenario \cite{inf}.
The possibility that some part of the energy density of the 
Universe is in a form other than particle-like matter 
has been envisaged, in particular in the form of
a constant energy ($\Lambda$CDM) \cite{lambda} or time-dependent
coherent energy density in a scalar field \cite{rp,decaying}.   
In this letter we discuss the cosmology of a model 
with a scalar field which has a simple exponential potential.
It is distinctly different from other scalar field cosmologies, 
in that its energy density plays a role from very early times, 
rather than just at recent epochs, and resembles much more
the `mixed dark matter' (MDM) model \cite{mdm} in which 
there is a component of matter which is collisionless 
during a period of the growth of structure.
The required potential has the merit that it arises
quite generically in particle theories involving 
compactifications such as supergravity or superstring theories,
and has (mainly for this reason) been quite extensively 
discussed in the context of inflationary models.

Let us first explain the properties of an exponential 
potential which make it a particular and interesting case.
The equations
of motion in an expanding FRW universe for the homogeneous 
mode of a scalar field $\phi$ with potential $V(\phi)$ 
coupled to ordinary matter only through gravity 
are 
\begin{eqnarray}
\ddot{\phi} + 2{\cal H} \dot{\phi} +
a^2V'(\phi)=\frac{1}{a^2}\frac{d}{d\tau}(a^2\dot{\phi}) + a^2V'(\phi)=0
\label{eq: potleoma}\\
{\cal H}^2 = \frac{1}{3M_{p}^2}(\frac{1}{2}\dot{\phi}^2 + a^2V(\phi)+a^2\rho_n) 
\label{eq: potleomb}\\
\dot{\rho_n} + n{\cal H} \rho_n = 0
\label{eq: potleomc}
\end{eqnarray}
where $\rho_n$ is the energy density in radiation ($n=4$)
or non-relativistic matter ($n=3$), ${\cal H}=\frac{\dot{a}}{a}$ 
is the conformal expansion rate of the universe with scale factor $a$,
dots are derivatives w.r.t. conformal time $\tau$, 
%(related to physical time $t$ by $t=\int a d\tau$), 
$'=\frac{d}{d\phi}$ and  $M_P=2.4 \times 10^{18}$GeV 
is the reduced Planck mass.
Multiplying (\ref{eq: potleoma}) by $\dot{\phi}$ and
integrating, one obtains     
\begin{eqnarray}
\rho_\phi(a) 
= \rho(a_o) e^{-\int_{a_o}^{a} 6(1 - \xi(a)) \frac 
{da}{a}}
\label{eq: scaling}
\end{eqnarray} 
where $\rho_\phi= \frac{1}{2a^2} \dot{\phi}^2+ V(\phi)$ is the total
scalar energy, and $\xi=V(\phi)/\rho_\phi$.  
In general therefore the energy density 
of a scalar field has the range of possible scaling 
behaviours $\rho \propto 1/a^m$
with $0 \leq m \leq 6$, and the scaling is completely determined
by the ratio of its potential to its kinetic energy.
%This statement is true independent of any specific 
%assumption about $\cal{H}$.

The special cosmological solutions in which we are interested here
are attractor solutions of (\ref{eq: potleoma}) - (\ref{eq: potleomc})
for the case of an exponential potential 
$V(\phi) = V_o e^{- \lambda{\phi/M_p}}$, which  were
given in \cite{rp} and \cite{wett}.
In these solutions the scalar field evolves so that
its total energy density $\rho_\phi$ scales in the same way as the
dominant component (i.e. $\rho_\phi  \propto 1/a^n$) and 
contributes a {\it fixed} fraction of the 
total energy density given by
\begin{equation}
\Omega_\phi\equiv \frac{\rho_\phi}{\rho_\phi + \rho_n}= \frac{n}{\lambda^2}
\qquad
\xi \equiv \frac{V(\phi)}{ \frac{1}{2a^2} \dot{\phi}^2+ V(\phi) }=1-\frac{n}{6} 
\label{attractor}
\end{equation}
for $\lambda > 1/\sqrt{n}$. Note that it is $\lambda$ alone
which determines the solution.
The existence of the attractor can be understood to follow from
the fulfillment of two conditions:
(i) $\rho_\phi$ scales {\it faster} than $1/a^n$ if $\rho_\phi >> \rho_n$
and, (ii) scales {\it slower} than $1/a^n$ if $\rho_\phi << \rho_n$.
These two behaviours tend to drive the two components to 
the attractor which lies between them.
That the first condition is satisfied can be seen from solving
(\ref{eq: potleoma}) - (\ref{eq: potleomc}) with $\rho_n=0$
for the exponential potential. There is then a
different set of attractors \cite{exp-attractor} in which 
\begin{equation}
\xi=1-\frac{\lambda^2}{6} \qquad \rho_\phi \propto \frac{1}{a^{\lambda^2}}
%\qquad \phi(t) \propto \frac{2}{\lambda} \ln (t) %\qquad a \propto t^A 
\label{eq: jhsoln}
\end{equation}
where $\lambda < \sqrt{6}$.% and $t=\int a(\tau)d\tau$. 
For $\lambda > \sqrt{6}$ there is
not a single attractor, but all solutions have $\xi \rightarrow 0$
asymptotically (and, therefore,  $\rho \propto 1/a^6$). 
The condition $\lambda > 1/\sqrt{n}$ for the attractor 
(\ref{attractor}) is indeed therefore just that anticipated. 
The second condition can be understood qualitatively as follows.
Taking, for simplicity,
the case $n=4$, (\ref{eq: potleoma}) - (\ref{eq: potleomc}) 
with $V(\phi)=0$ give $\dot\phi(t)= \dot\phi_o (\frac{a_o}{a})^2$ 
and therefore
\begin{eqnarray}
\phi(\tau)= \phi_o +\dot\phi_o \tau_o\ln \frac{\tau}{\tau_o} 
\qquad \rho_\phi >> \rho_{\gamma}\\
\phi(\tau)= \phi_o +\dot\phi_o \tau_o
\big( 1- \frac {\tau_o}{\tau}\big)
\qquad \rho_\phi << \rho_{\gamma}
\label{eq: phiradn}
\end{eqnarray}
These will also hold as approximate solutions in the case that 
the potential energy is sub-dominant. The first solution shows 
how, for a sufficiently steep exponential, the potential energy 
can remain small relative to the kinetic energy
($\sim 1/\tau^3$) so that the rapid scaling (associated with $\xi << 1$)
can be maintained. The field in both the attractor solutions 
(\ref{attractor}) and (\ref{eq: jhsoln}) has this same logarithmic
time dependence. On the other hand, the second limit shows how 
the larger damping due to radiation domination slows down the
evolution of the field giving an almost constant potential energy 
which will thus ultimately catch up with the kinetic energy, increasing 
$\xi$ and causing the scalar energy to scale slower.

A potential which is less steep than the exponential will not 
satisfy the first condition \cite{footnote-on-oscillation}, and a 
steeper potential (e.g. $\sim e^{-\phi^2/M_P^2}$) will always 
decay asymptotically relative to the other components. The
existence of this particular attractor cosmological solution 
is thus quite specific to the exponential potential. Further
this is in fact a potential which can arise quite generically 
in particle physics theories involving compactified dimensions 
(with internal dimensions characterized by $M_P$). For this reason 
it has been considered quite extensively in the context 
of inflation \cite{exp-attractor,exp-inflation1,exp-inflation2}, 
since for $\lambda < \sqrt{2}$ the solutions (\ref{eq: jhsoln})
describe `power-law' inflation (with  $a \propto t^{2/\lambda^2}$
in terms of physical time  $t=\int a d\tau$). 
Examples of specific supergravity theories in which such potentials 
are obtained are given in \cite{exp-inflation1}, and various higher
dimensional theories of gravity in which they arise discussed in 
detail in \cite{exp-inflation2}  and \cite{halliwell-exp}.

If such a field does exist, it will enter the attractor and
contribute a fraction of the energy density (fixed by $\lambda$)
at some time determined by its initial energy density.
Nucleosynthesis provides the earliest constraint on how
large such a contribution can be.  The expansion rate of the 
Universe at nucleosynthesis is 
increased over its standard model value by the same amount as
$\Delta N_{eff}$ relativistic degrees of freedom, with 
$\Omega_\phi=\frac{3}{4}\frac{7\Delta N_{eff}/4}{10.75+7\Delta N_{eff}/4}$,
where $\Omega_\phi$ is the fraction contributed in the matter era.
There is some disagreement on the precise nucleosynthesis constraint
on $\Delta N_{eff}$, but a bound of 
$\Delta N_{eff}=0.9$ is given by various authors \cite{cst} or 
even a more conservative one of $\Delta N_{eff}=1.5$ by others
\cite{ks}, which corresponds to $\Omega_\phi<0.1-0.15$.

{\it Prima facie} this constraint would seem to require 
entry into the attractor after nucleosynthesis if the scalar 
field is to play any significant role cosmologically \cite{rp}.
The requirement of entry after nucleosynthesis would apparently
mandate the unattractive fine-tuning 
(typical of scalar field models) of the initial energy density
in the potential to some small value.  It was in fact the
incorrectness of this second assumption
which motivated the present study: If, prior to
nucleosynthesis, the energy density in the exponential field
with  $\lambda > \sqrt{6}$ dominates over that in the radiation,
there will typically be a long transient period after 
$\rho_\phi \sim \rho_\gamma$ during which the scalar energy is very
sub-dominant (much less than its value in the attractor (\ref{attractor})). 
This is simply because the ratio $\xi \rightarrow 0$ 
in the kinetic energy dominated pure scalar cosmology, but is of order 
one in the attractor with radiation. During the time that
$\xi$ is increasing (potentially many expansion times as it
cannot grow faster than $a^6$) the scalar field energy continues
to red-shift away as $1/a^6$.  Such a dominance by kinetic
energy can occur in certain post-inflationary cosmologies
which have considerable interest in their own right 
\cite{mj},\cite{spokoiny}. 
It has transpired from the present work however 
that the first reason for disregarding this model
is also incorrect, and that entry to the attractor prior
to nucleosynthesis is in fact consistent - quite simply because 
the small contribution has a compensating long time to act. 

\begin{figure}
\centerline{\psfig{file=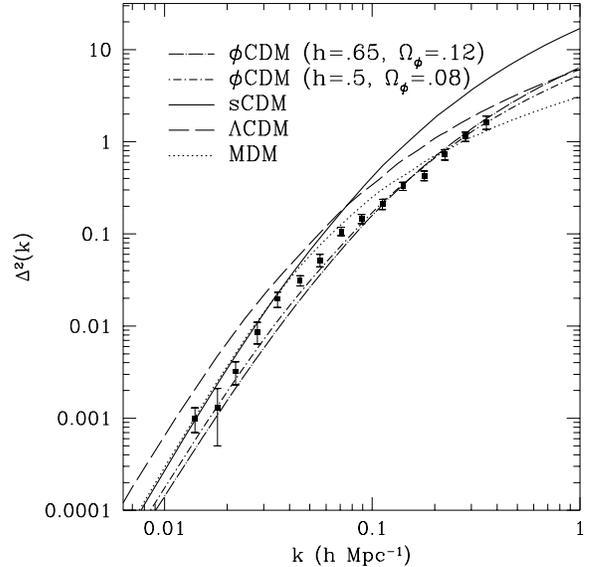,width=3.2in}}
\caption{Mass variance per unit $\ln k$ computed from Boltzman
code for different models compared with that inferred from a compilation of
galaxy surveys \protect\cite{PD}}
\label{fig1}
\end{figure}

\begin{figure}
\centerline{\psfig{file=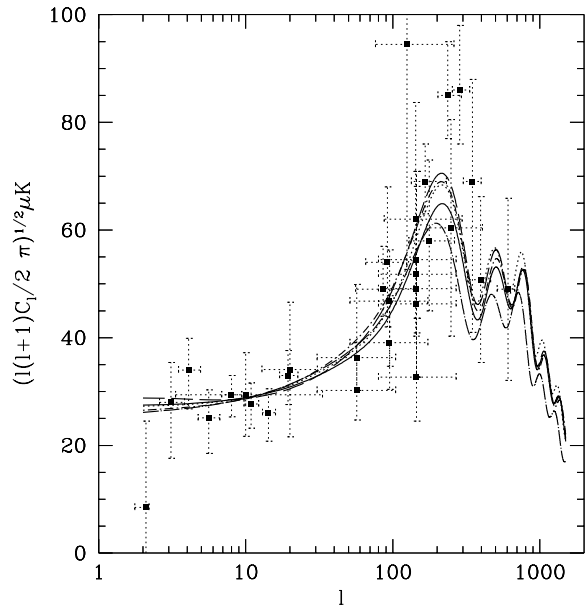,width=3.4in}}
\caption{Comparison of different model predictions to current
experimental data. All models were COBE normalized and are labeled
as in Figure 1.}
\label{fig2}
\end{figure}

We have carried out a detailed calculation of the evolution
of perturbations in this cosmology (which we refer to as $\phi$CDM).
We assume that the attractor is established at the beginning
of our numerical simulation, deep in the radiation era, and take 
an initial standard inflationary scale-invariant spectrum of 
adiabatic perturbations. The relevant equations are
the linearized  coupled Einstein-Boltzmann equations given in 
\cite{pert}, supplemented by the scalar field and its perturbations
$\phi_{total}=\phi(\tau)+\varphi(\tau,{\bf x})$, with
evolution equation 
\begin{eqnarray}
{\ddot \varphi}&+&2{\cal H}{\dot
\varphi}-{\nabla^2}\varphi+a^2V''\varphi
+{1 \over 2}{\dot \phi}{\dot \gamma}=0 \label{sfev}
\end{eqnarray}
and additional components to the perturbed energy-momentum tensor:
\begin{eqnarray}
a^2\delta T^0_0 &=&-{\dot \phi}{\dot \varphi}-a^2V'\varphi \nonumber \\
-a^2\partial_i\delta T^0_i &=&{\dot \phi}\nabla^2\varphi \nonumber \\
a^2\delta T^i_i&=&3{\dot \phi}{\dot \varphi}-3a^2V'\varphi
\end{eqnarray}
where $\gamma$ is the trace of the metric perturbation. 
We vary $\Omega_\phi$ and $h$ (where $H_0=h 100$km/s/Mpc
is the Hubble constant today), keeping the remaining 
cosmological parameters fixed at the values of sCDM, and 
find the best fit model to both CMB and large scale structure. 
To do this we use the COBE measurement of CMB anisotropies on 
large scales \cite{smoot} to normalize our theory \cite{BW}, estimate the
theoretical mass variance per unit $\ln k$, $\Delta^2(k)$ and compare with
that rendered from a collection of galaxy surveys \cite{PD}.
In Figure \ref{fig1} we show $\Delta^2(k)$ for two best fit  
$\phi$CDM models, for sCDM, for a $\Lambda$CDM universe with
$\Omega_\Lambda=0.6$, and for an MDM model with $\Omega_\nu=0.2$ 
in the form of two massive neutrinos species. 
It is clear that for these values $\phi$CDM fares as well or better
than the other models.
Another useful quantity to work with is the mass fluctuations
on $8$h$^{-1}$Mpc scales, $\sigma_8^2=\int_0^\infty\frac{dk}{k}\Delta^2(k)
(\frac{3j_1(kR)}{kR})^2\mid_{R=8}$.
This can be related to masses and abundances of rich clusters and
supplies us with a very tight constraint on possible cosmologies;
indeed current estimates give $\sigma_8=0.6\pm0.1$ \cite{cluster}.
A good fit to $\sigma_8$ is
\begin{eqnarray}
\sigma_8(\Omega_\phi)=e^{-8.7\Omega_\phi^{1.15}}\sigma^{CDM}_8
\end{eqnarray}
where $\sigma^{CDM}_8$ is the COBE normalized sCDM $\sigma_8$.
Again we see that there is range of values of $\Omega_\phi$ and 
$H_0$ which satisfy the above constraint {\it and} are consistent
with the limits imposed by BBN. 
In Figure \ref{fig2} we compare the $C_\ell$s of our models
with a compilation of data points \cite{exp}. Again they are
consistent with the current data.

The evolution of perturbations in the presence of
the scalar field is simple to understand.
On superhorizon scales there is the usual growing mode with
$\delta_c$,$\varphi$ $\propto \tau^2$ (where $\delta_c$ is the 
density contrast in the CDM). This is to be expected; the 
superhorizon evolution is insensitive to the ``chemistry'' of 
the matter and totally dominated by gravity. On sub-horizon scales in the
radiation era, the Meszaros effect comes into play giving
$\delta_c\propto \ln \tau$. The specific effect of the scalar
field appears on subhorizon scales in the matter era. 
The perturbation in the scalar field itself has the approximate 
solution $\varphi\propto\frac{1}{\tau^{3/2}}J_{\frac{3}{2}}(k\tau)$ (where
$J_\nu$ is a Bessel function) which, when fed back into the equation
for $\delta_c$ gives an altered solution for the usual
growing mode $\delta_c \propto \tau^{2-\epsilon}$ where
\begin{eqnarray}
\epsilon=\frac{5}{2}(1-
\sqrt{1-\frac{24}{25}\Omega_{\phi}}) \label{supp}
\end{eqnarray}
This solution shows explicitly how even a small contribution 
from the scalar field can give a significant effect, as it
acts all the way through the matter era.
The expected suppression of $|\delta_c|^2$  for  modes
larger than $k_{eq}$ is of order $(1+z_{eq})^{-\epsilon}$, where $k_{eq}$
is the wavenumber of the horizon size at radiation-matter equality. 
This last effect is reminiscent of the evolution of perturbations
in a mixed dark matter (MDM) universe where one has component
of matter, $\rho_\nu$ which is collisionless for a period of time during the
matter era \cite{BES}.

It is useful to pursue a comparison between $\phi$CDM and MDM 
to identify the key differences.
Firstly the scaling behaviour of the additional background energy density
differs: While for $\phi$CDM the energy density
in $\phi$ follows the dominant form of energy quite closely,
for MDM  $\rho_{\nu}$ changes from scaling as $1/a^{4}$ to 
scaling as $1/a^{3}$  when $3k_BT_{\nu}
\simeq m_{\nu}$ where $T_{\nu}$
($m_{\nu}$) is the massive neutrino temperature (mass) and
$k_B$ is the Boltzmann constant. For a period  between
matter-radiation equality and this transition $\Omega_{\nu}$
is smaller than its asymptotic value, and there is 
less suppression of growth in the CDM than in the case of 
the scalar field.
A further difference is that the period of time during which
perturbations are suppressed is shorter in MDM compared
to $\phi$CDM. In both cases there is a wavenumber $k_{su}$ 
which separates growing modes from damped modes. For $\phi$CDM 
this scale is roughly the horizon i.e. $k_{su}\propto \frac{1}{\tau}$, 
while for MDM it is the free streaming scale i.e. 
$k_{su}=8a^{1/2}(m_\nu/10eV)h$Mpc$^{-1}\propto \tau$. 
Clearly in the latter case any given mode of $\delta_c$ will 
eventually start to grow. In particular modes around $k_{eq}$
will already have started to undergo collapse.
A final important difference concerns the evolution of
perturbations in the radiation era. For MDM, the perturbation in the
massive neutrinos behaves much like radiation until it is well inside
the horizon, and  this transition is set by the Jeans scale i.e.
when $k\tau\simeq 1/c_s=\sqrt{3}$. For the scalar field, on the
other hand, the transition occurs for larger wavelengths, $k\tau<1$.
This means that perturbations in the CDM will stop growing earlier 
in $\phi$CDM than in MDM.
The accumulated effect of these differences explains what we
have observed - that,  with half the energy density
of MDM with two massive neutrinos, $\phi$CDM brings about 
approximately the same suppression of power on small scales.
The fact that this suppression lasts until today
leads to the formation of structure at higher redshifts.
There is now strong observational evidence that any
model of structure formation must have this feature. 

Let us now turn to the effect that the scalar field has on
the CMB. We shall rely on
the simplified picture of \cite{HSS} to understand
the angular power spectrum, $C_\ell$,
defined
as $C(\theta)=\langle\frac{\Delta T}{T}({\bf n})\frac{\Delta T}{T}({\bf
n'})\rangle= (4\pi)^{-1}\sum(2\ell+1)C_\ell P_\ell(\cos\theta)$,
with ${\bf n}\cdot{\bf n'}=\cos\theta$.
For $\ell>100$ the main features of the $C_\ell$s
are given by the power spectrum of radiation perturbations
at last scattering, $\delta_\gamma$. Ignoring projection effects,
one has that the structure  of the peaks and troughs are
given by
\begin{eqnarray}
<|\delta_\gamma|^2>\propto\cos^2(kr_s) \ \ \ \ kr_s>1
\end{eqnarray}
  where $r_s$ is the sound horizon in the 
baryon-photon fluid, $r_s=\int_0^{\tau_*}
\frac{d\tau}{3[1+R(\tau)]}$ and $R=\frac{3\rho_B}{4\rho_\gamma}$.
The spatial frequency $k$ is roughly related to the angular frequency
$\ell$. The fact that the properties of the $C_\ell$s are 
dominated by this quantity at $a\simeq10^{-3}$ means 
that the effect of $\phi$ on the CMB will be much 
smaller than its net effect on $\delta_c$. 
Adding the scalar field
component brings about two effects which we can understand 
qualitatively. Firstly the oscillations are shifted to higher $\ell$s. 
Because of the additional energy density in the scalar field,
the expansion rate will be larger and the conformal horizon will be 
smaller for the same red-shift in $\phi$CDM compared sCDM.
This feeds through to give a different $r_s$ for the same value of $a$,
shifting the peaks as observed. 
The other main feature is an increase in power 
in the peaks. This can be understood easily
using the picture outlined in \cite{HSS}. The oscillations
in $\delta_\gamma$ are driven by the evolution in the gravitational
potentials, and here as in the MDM case \cite{DGS} the
change in the growth of metric perturbations
boosts the amplitude of the peaks by a few percent.

We conclude that the cosmological model we have studied provides an
interesting and distinct alternative to other models
which have been proposed. It has the attractive feature that
$\lambda$ ($=\sqrt{3/\Omega_\phi}$), the single extra parameter 
compared to standard CDM, has a value which is of the order 
naturally expected in the many particle physics theories 
in which the field arises.  
With the launch of high resolution 
space based experiments, such as the Planck explorer and the 
MAP satellite, it should be possible to distinguish the effect 
on the CMB of such an exponential scalar field if it exists, or to
rule out its existence and place tighter constraints 
on the physical theories in which these fields arise
\cite{CP}. 

{\it Acknowledgements}: 
We acknowledge U. Seljak and
   M. Zaldarriaga for the use of their Boltzmann-Einstein solver.
 We thank A. Albrecht, C. Balland, M. Davis, A. Jaffe, J. Levin, A. Liddle
 and J. Silk for conversations.  P.F.
is supported by the Center for Particle Astrophysics, a NSF Science and
Technology Center at U.C.~Berkeley, and MJ by an Irish Government
(Dept. of Education) post-doctoral fellowship.

%Correspondance should be addressed to {\tt pgf@physics.berkeley.edu}
%\eject 
%\pagestyle{empty}

\end{document}